\documentclass[aps,pra,reprint,showpacs,groupedaddress]{revtex4-1}
\usepackage{amsmath,amssymb,mathrsfs,amsfonts}
\usepackage{epsfig}
\usepackage{bm}
\usepackage{color}
\usepackage[dvipdfm,
            pdfstartview=FitH,
            CJKbookmarks=true,
            bookmarksnumbered=true,
            bookmarksopen=true,
            linktocpage=true,
            colorlinks=true, 
            pdfborder=001,   
            citecolor=blue,  
            urlcolor=blue,   
            linkcolor=blue,  
            anchorcolor=blue,
            ]{hyperref}      

\begin{document}
  \title{Quantum dephasing induced by non-Markovian random telegraph noise}
  \author{Xiangji Cai}
   \email{xiangjicai@foxmail.com}
  \affiliation{School of Science, Shandong Jianzhu University, Jinan 250101, China}

\begin{abstract}
  We theoretically study the dynamical dephasing of a quantum two level system interacting with an environment which exhibits non-Markovian random telegraph fluctuations.
  The time evolution of the conditional probability of the environmental noise is governed by a generalized master equation depending on the environmental memory effect.
  The expression of the dephasing factor is derived exactly which is closely associated with the memory kernel in the generalized master equation for the conditional probability of the environmental noise.
  In terms of three important types memory kernels, we discuss the quantum dephasing dynamics of the system and the non-Markovian character exhibiting in the dynamical dephasing induced by non-Markovian random telegraph noise.
  We show that the dynamical dephasing of the quantum system does not always exhibit non-Markovian character which results from that the non-Markovian character in the dephasing dynamics depends both on the environmental non-Markovian character and the interaction between the system and environment.
  In addition, the dynamical dephasing of the quantum system can be modulated by the external modulation frequency of the environment.
  This result is significant to quantum information processing and helpful for further understanding non-Markovian dynamics of open quantum systems.
\end{abstract}

\maketitle

\section{Introduction}
\label{sec:intr}

Quantum coherence plays an important role in quantum mechanics and has recently received much attention with the development of experimental technique
to observe and control quantum systems~\cite{PhysRevLett.113.140401,RevModPhys.89.041003,PhysRep.762.1}.
A realistic quantum system always unavoidably interacts with the environment,
which leads to decoherence during its dynamical evolution.
In the past decades, the decoherence dynamics of open quantum systems has been generally studied
within the framework of Markov approximation.
Investigations on the decoherence dynamics of open quantum systems beyond Markov approximation
has increasingly drawn much attention in a wide variety of fields ranging from understanding the basic features of quantum mechanics to the applications of advanced experimental technique of coherent manipulation
and control in quantum information science, such as, the quantification and measurement of the non-Markovian character in
the quantum dynamics~\cite{Breuerbook,PhysRevLett.103.210401,PhysRevLett.105.050403,PhysRevLett.112.120404,RepProgPhys.77.094001,
SciRep.4.5720,PhysRevA84.031602,PhysRevLett.112.210402,PhysRevA90.052118,RevModPhys.88.021002,RevModPhys.89.015001}.
It has been demonstrated both theoretically and experimentally that non-Markovian dynamical decoherence gives rise to incomplete loss of coherence due to the backaction of coherence from the environment, which makes the quantum system have long correlation time with the environment.

Many theoretical methods have been well established to investigate the non-Markovian character in the dynamical decoherence of open quantum systems, such as, non-Markovian quantum state diffusion~\cite{PhysRevA58.1699,PhysRevA.60.91,PhysRevLett.82.1801}, projection operator~\cite{PhysRevA59.1633,PhysRevB70.045323,PhysRevE72.056106,PhysRevE73.016139,Entropy21.1040}, quantum jumps~\cite{PhysRevA79.062112}, nonequilibrium Green-function~\cite{PhysRevB78.235311,PhysRevLett.109.170402} and dynamical maps~\cite{PhysRevA71.020101,PhysRevA87.030101}.
Non-Markovian dynamical decoherence induced by the interaction of a quantum system with its
environment can be also accurately modeled by means of stochastic processes with fixed
statistical properties of the environmental noise within the framework of Kubo-Anderson spectral diffusion.
There have been well-established theoretical investigations on the decoherence
dynamics of open quantum systems which usually assume that the environmental noise
with Markovian statistical properties based on classical and quantum treatments~\cite{JPhysSocJpn.9.316,JPhysSocJpn.9.935,
PhysRevLett.88.228304,PhysRevLett.94.167002,PhysRevLett.118.140403,PhysRevA95.052126,PhysRevA97.042126,JChemPhys.133.241101,JPhysB45.154008,JChemPhys.139.024109,RevModPhys.86.361,JChemPhys.142.094106,
JChemPhys.142.094107,Entropy17.2341,JStatMech.2016.054016,JChemPhys.148.014103,JChemPhys.148.014104,IntJTheorPhys.57.1082}.
Random telegraph noise (RTN), as an important non-Gaussian noise,
has been widely used to theoretically model the environmental effects on quantum systems
in a large variety of fundamental physical, chemical and biological processes,
such as, fluorescence process of single molecules~\cite{PhysRevLett.90.238305,AnnuRevPhysChem.55.457,PhysRevLett.90.120601}, rate process in chemical reactions~\cite{RevModPhys.62.251}, entanglement and decoherence processes induced by low-frequency $1/f^{\alpha}$ noise~\cite{PhysRevB79.125317,PhysRevB90.054304,PhysRevB94.235433} and frequency modulation process in quantum information science~\cite{RepProgPhys.80.056002}.
Moreover, it has been well-established experimental investigations on the dephasing dynamics of open quantum systems driven by a classical stochastically fluctuating field exhibiting non-Markovian random telegraph fluctuations~\cite{ApplPhysLett.110.081107,PhysRevA100.052104}.

There are many important physical situations where a quantum system may be coupled to a composite
or structured environment.
In these situations, due to the essential role of the couplings between the sub-environments, both the dynamical evolution of the quantum system and the environmental statistical properties display dominant non-Markovian characters,
such as, cavity quantum electrodynamics and quantum measurement processes~\cite{PhysRevLett.106.233601,PhysRevB77.075325,PhysRevA90.042108,PhysRevA91.042111,PhysRevA92.012315,PhysRevA95.023610,PhysRevB96.235417}.
Hence, it is necessary to take into extensive consideration the memory effects of the environment, namely, the non-Markovian statistical properties of the environmental noise, to study the dynamical evolution of the quantum system.
Recently, the non-Markovian RTN with an exponentially correlated memory kernel has been discussed~\cite{PhysRevE50.2668}
and widely used to study the relevant issues of open quantum systems~\cite{PhysRevA94.042110,PhysRevA95.052104,EurophysLett.118.60002,JChemPhys.149.094107,EurophysLett.125.30007}.

In this paper, we theoretically study the dephasing dynamics of a quantum two level system which interacts with a noisy environment exhibiting non-Markovian RTN statistical properties.
The non-Markovian character of the environmental noise is governed by a generalized master equation for the time evolution of the conditional probability.
Based on the environmental non-Markovian character, the exact expression of the dephasing factor is derived which is closely associated with the memory kernel of the environmental noise.
In the presence of three important types memory kernels, we discuss the quantum dephasing dynamics of the system and the non-Markovian character in the dynamical dephasing induced by the non-Markovian RTN.
We show that the non-Markovian character exhibiting in the quantum dephasing dynamics depends on both the environmental non-Markovian character and the interaction between the system and environment, and that the dynamical dephasing of the quantum system does not always exhibit non-Markovian character.
Moreover, we show that the quantum dynamical dephasing of the system can be modulated by the external modulation frequency of the environment.

\section{Dynamical dephasing in a noisy environment}
We consider the quantum dynamical dephasing of a two level system in the presence of a noisy environment which exhibits non-Markovian character.
The environmental effects only influence the coherence of the quantum system and the energy of the system is conserved.
Based on the Kubo-Anderson spectral diffusion model, the environmental influences on the quantum system lead to that the transition frequency fluctuates stochastically as~\cite{JPhysSocJpn.9.316,JPhysSocJpn.9.935}
\begin{equation}
\label{eq:totrhoge}
\omega(t)=\omega_{0}+\xi(t),
\end{equation}
where the intrinsic transition frequency is denoted by $\omega_{0}$ in the basis $\{|e\rangle,|g\rangle\}$ and the stochastic fluctuation term $\xi(t)$ obeys a stationary stochastic process caused by the environmental noise.

The dynamical evolution of the off-diagonal element of the total density matrix in the basis $\{|e\rangle,|g\rangle\}$ satisfies the stochastic differential equation~\cite{PhysRevA95.052104}
\begin{equation}
\label{eq:totrhoge}
\frac{d}{dt}\rho_{ge}\bm(t;\xi(t)\bm)=i[\omega_{0}+\xi(t)]\rho_{ge}\bm(t;\xi(t)\bm),
\end{equation}
where the notation $\rho_{ge}\bm(t;\xi(t)\bm)$ is employed to indicate the dependence of the environmental noise $\xi(t)$.
By iterating Eq.~\eqref{eq:totrhoge} and taking an average over the environmental noise $\xi(t)$, we obtain the off-diagonal element of the reduced density matrix
\begin{equation}
\label{eq:redrhoge}
\rho_{ge}(t)=\langle\rho_{ge}\bm(t;\xi(t)\bm)\rangle=e^{i\omega_{0}t}F(t)\rho_{ge}(0),
\end{equation}
with $\langle\cdots\rangle$ denoting an average of different realizations over the environmental noise $\xi(t)$.
Here it has been assumed that the system and environment are initially independent
\begin{equation}
\label{eq:averhoge}
  \rho_{ge}(0)=\left\langle\rho_{ge}\bm(0;\xi(0)\bm)\right\rangle=\rho_{ge}\bm(0;\xi(0)\bm),
\end{equation}
and $F(t)$ is the dephasing factor quantifying the coherence evolution of the quantum system which is defined as the Dyson series expansion in terms of the moments of the environmental noise~\cite{PhysRevA94.042110,JChemPhys.149.094107}
\begin{equation}
\label{eq:depfac}
\begin{split}
F(t)&=\left\langle\exp\left[{i\int_{0}^{t}dt'\xi(t')}\right]\right\rangle\\
    &=1+\sum_{n=1}^{\infty}i^{n}\int_{0}^{t}dt_{1}\cdots\int_{0}^{t_{n-1}}dt_{n}\langle\xi(t_{1})\cdots\xi(t_{n})\rangle.
\end{split}
\end{equation}
Based on the stationary character of the environmental noise, the odd moments of $\xi(t)$ vanish and the time dependent dephasing factor $F(t)$ in Eq.~\eqref{eq:depfac1} is real-valued.

To further describe the dynamical dephasing, we employ two physical quantities in terms of the dephasing factor $F(t)$:
One is the dephasing rate~\cite{EurophysLett.125.30007}
\begin{equation}
\label{eq:deprat}
\gamma(t)=-\frac{\dot{F}(t)}{F(t)}=-\frac{1}{|F(t)|}\frac{d}{dt}|F(t)|,
\end{equation}
and another is the non-Markovianity~\cite{RevModPhys.88.021002}
\begin{equation}
\label{eq:non-Mar}
 \mathcal{N}=-\int_{0 \ \gamma(t)<0}^{\infty}\gamma(t)|F(t)|dt,
\end{equation}
which are closely related to the exchange of a flow of coherence information between the quantum system and environment~\cite{PhysRevLett.103.210401,PhysRevA81.062115}:
When $\gamma(t)$ is negative in some time interval, the quantum system begins to gain the information flowing from the environment and it indicates that the dephasing dynamics of the quantum system exhibits some non-Markovian character due to the coherent backaction from the environment.
When $\gamma(t)$ always takes positive values, the information flows continuously and unidirectionally from the system into the environment and it indicates that the dephasing dynamics of the quantum system only exhibits Markovian character due to no backaction of coherence from the environment.
The non-Markovianity $\mathcal{N}$ denotes the maximal total amount of the backflow of information from the environment to the system in the dynamical evolution.

\section*{Non-Markovian RTN}

We assume that the environmental noise $\xi(t)$ obeys a stationary non-Markovian RTN process
which jumps randomly between the values $\pm1$ with with an amplitude $\nu$ and an average transition rate $\lambda$.
For this noise process, the time evolution of the conditional probability is governed by a generalized master equation~\cite{ZPhysB26.85,PhysRep.88.207,PhysRevE50.2668}
\begin{equation}
\label{eq:conproevo}
\begin{split}
  \frac{\partial}{\partial t}P(\nu,t|\xi',t')=&\int_{t'}^{t}d\tau K(t-\tau)\big[-\lambda P(\nu,\tau|\xi',t')\\
                                              &+\lambda P(-\nu,\tau|\xi',t')\big],\\
  \frac{\partial}{\partial t}P(-\nu,t|\xi',t')=&\int_{t'}^{t}d\tau K(t-\tau)\big[-\lambda P(-\nu,\tau|\xi',t')\\
                                               &+\lambda P(\nu,\tau|\xi',t')\big],
\end{split}
\end{equation}
where $K(t-\tau)$ denotes the memory kernel of the environmental noise and the initial condition is $P(\xi,t'|\xi',t')=\delta_{\xi,\xi'}$ for $\xi=\pm\nu$.
The environmental noise $\xi(t)$ is non-Markovian because the Chapman-Kolmogorov equation is no longer valid unless the memory kernel is proportional to a $\delta$ function~\cite{Kampenbook}.
The solution of the conditional probability can be obtained by taking the Laplace transform over Eq.~\eqref{eq:conproevo} as $\widetilde{P}(\xi,p|\xi',t')=\int_{0}^{+\infty}P(\xi,t|\xi',t')e^{-pt}dt$.
In Laplace domain, the conditional probability can be analytically solved as
\begin{equation}
\label{eq:Lapconpro}
\begin{split}
\widetilde{P}(\xi,p|\xi',t')=&\frac{p+\lambda\widetilde{K}(p)}{p[p+2\lambda\widetilde{K}(p)]}\delta_{\xi,\xi'}
+\frac{\lambda\widetilde{K}(p)}{p[p+2\lambda\widetilde{K}(p)]}\delta_{\xi,-\xi'}\\
=&\frac{1}{2}\left[\frac{1}{p}+\frac{1}{p+2\lambda\widetilde{K}(p)}\right]\delta_{\xi,\xi'}\\
&+\frac{1}{2}\left[\frac{1}{p}-\frac{1}{p+2\lambda\widetilde{K}(p)}\right]\delta_{\xi,-\xi'},
\end{split}
\end{equation}
where $\widetilde{K}(p)$ is the Laplace transform of the memory kernel of the environmental noise.
In Eq.~\eqref{eq:Lapconpro}, we derive the exact expression of the conditional probability of a non-Markovian RTN process with an arbitrary memory kernel $K(t-\tau)$ in the Laplace domain.
To obtain the solution of the time domain conditional probability $P(\xi,t|\xi',t')$, we should take the inverse Laplace transform over Eq.~\eqref{eq:Lapconpro}. In the memoryless limit of the environmental noise, namely, the memory kernel $K(t-\tau)=\delta(t-\tau)$ and the statistical property of the environment only exhibiting Markovian character, the conditional probability in time domain can be written as $P(\xi,t|\xi',t')=\frac{1}{2}[1+e^{-2\lambda(t-t')}]\delta_{\xi,\xi'}+\frac{1}{2}[1-e^{-2\lambda(t-t')}]\delta_{\xi,-\xi'}$, which returns to the previous result given in Ref.~\onlinecite{Kampenbook}.
Due to the stationary character of the environmental noise, the one-point unconditional probability is time independent
\begin{equation}
\label{eq:onepro}
  P(\xi,t)=\int P(\xi,t|\xi_{0},0)P(\xi_{0})d\xi_{0}=\frac{1}{2}\delta_{\xi,\nu}+\frac{1}{2}\delta_{\xi,-\nu},
\end{equation}
where the initial distribution is stationary $P(\xi_{0})=\frac{1}{2}\delta_{\xi_{0},\nu}+\frac{1}{2}\delta_{\xi_{0},-\nu}$.

We focus on the statistical characteristics of the non-Markovian RTN. The noise process is with zero average
\begin{equation}
\label{eq:ave}
  \langle\xi(t)\rangle=0,
\end{equation}
and its second-order correlation function satisfies
\begin{equation}
\label{eq:Lapcor}
  C(t-t')=\langle\xi(t)\xi(t')\rangle=\mathcal{L}^{-1}[\widetilde{C}(p)],
  \quad\widetilde{C}(p)=\frac{\nu^{2}}{p+2\lambda\widetilde{K}(p)}.
\end{equation}
Here $\mathcal{L}^{-1}$ denotes the inverse Laplace transform.
It is clear that the second-order correlation function of a non-Markovian RTN process satisfying the generalized master equation~\eqref{eq:conproevo} closely depends on the memory kernel of the environmental noise.
Based on the Bayes rule, the higher odd-order correlation functions vanish
\begin{equation}
\label{eq:oddcor}
  \langle\xi(t_{1})\xi(t_{2})\cdots\xi(t_{2n-1})\rangle=0,
\end{equation}
and the higher even-order correlation functions factorize as
\begin{equation}
\label{eq:evencor}
  \langle\xi(t_{1})\xi(t_{2})\cdots\xi(t_{2n})\rangle=\langle\xi(t_{1})\xi(t_{2})\rangle\langle\xi(t_{3})\cdots\xi(t_{2n})\rangle,
\end{equation}
for all sets of time sequences $t_{1}>t_{2}>\cdots>t_{2n-1}>t_{2n}$ ($n\geq2$).
The factorization relations in Eqs.~\eqref{eq:oddcor} and~\eqref{eq:evencor} for the higher-order correlation functions of a stationary non-Markovian RTN process were first derived in Ref.~\onlinecite{PhysRevE50.2668} which were also derived when a stationary RTN process exhibits only Markovian character~\cite{Physica65.303,PhysRevA30.2390}. It is worth noting that the more general factorization relations for the higher-order correlation functions of the environmental noise $\langle\xi(t)\cdot\cdot\cdot\xi(t_{n})\rangle=\langle\xi(t)\xi(t_{1})\rangle\langle\xi(t_{2})\cdot\cdot\cdot\xi(t_{n})\rangle$
for $t>t_{1}>\cdot\cdot\cdot>t_{n}$ $(n\geq2)$ when the RTN process exhibits nonstationary character and in the limit of stationary character of the environmental noise, it recovers to the factorization relations for the higher odd- and even-order correlation functions in Eqs.~\eqref{eq:oddcor} and~\eqref{eq:evencor} due to the vanishing of the odd moments of the environmental noise.

\section*{Exact solution of the dephasing factor}

Based on the statistical characteristics of the environmental noise $\xi(t)$ obtained above, we can derive that the partial cumulants higher than the second order vanish
\begin{equation}
\label{eq:avecor}
  \langle\xi(t)\xi(t_{1})\cdots\xi(t_{n})\rangle^{\mathrm{pc}}=0,\\
\end{equation}
for every ordered set of time instants $t>t_{1}>\cdots>t_{n} (n\geq2)$.
Consequently, the time evolution of the dephasing factor yields the integro-differential equation as follows~\cite{JChemPhys.149.094107}
\begin{equation}
\label{eq:intdif}
  \frac{d}{dt}F(t)=-\int_{0}^{t}C(t-t')F(t')dt',
\end{equation}
with the initial condition $F(0)=1$.
By means of Laplace transform of Eq.~\eqref{eq:intdif} and based on the second-order correlation function in Eq.~\eqref{eq:Lapcor}, the dephasing factor can be analytically expressed as
\begin{equation}
\label{eq:depfacexp}
   F(t)= \mathcal{L}^{-1}\big[\widetilde{F}(p)\big],\quad\widetilde{F}(p) = \frac{p+2\lambda\widetilde{K}(p)}{p\big[p+2\lambda\widetilde{K}(p)\big]+\nu^{2}}.
\end{equation}
It is clear that the dephasing factor induced by a non-Markovian RTN process is closely associated with the memory effect  of the environmental noise.
Here, Eq.~\eqref{eq:depfacexp} gives the exact expression of the dephasing factor induced by a non-Markovian RTN process with an arbitrary memory kernel $K(t-\tau)$ in Laplace domain.
In contrast to our previous work in Refs.~\onlinecite{PhysRevA94.042110,JChemPhys.149.094107},
it is more convenient to calculate the dephasing factor induced by a non-Markovian RTN process based on the expression derived in Eq.~\eqref{eq:depfacexp} rather than to calculate the statistical characteristics of the environmental noise in terms of Eqs.~\eqref{eq:conproevo}-~\eqref{eq:evencor} and then to derive the closed differential equation for the dephasing factor based on the closure of the higher-order correlation functions of the environmental noise~\cite{PhysRevA94.042110,JChemPhys.149.094107}.

To derive the exact expression of the dephasing factor $F(t)$ in time domain, we can take inverse Laplace transform over the Laplace domain dephasing factor $\widetilde{F}(p)$ in Eq.~\eqref{eq:depfacexp}.
The Laplace domain dephasing factor $\widetilde{F}(p)$ in Eq.~\eqref{eq:depfacexp} can be generally expressed as a proper rational function which is the quotient of two real polynomials
\begin{equation}
\label{eq:polynomials}
   \widetilde{F}(p) = \frac{p+2\lambda\widetilde{K}(p)}{p\big[p+2\lambda\widetilde{K}(p)\big]+\nu^{2}}=\frac{N(p)}{D(p)},
\end{equation}
where the degrees of the numerator polynomial $N(p)$ and denominator polynomial $D(p)$ are $m$ and $n$ ($m<n$), respectively.
The exact expressions of the numerator polynomial $N(p)$ and denominator polynomial $D(p)$ are closely associated with the Laplace transform of the memory kernel of the environmental noise.
To derive the time domain dephasing factor, we should obtain the roots of the denominator polynomial $D(p)$.
For the case that the degree of $D(p)$ is two or three (quadratic or cubic polynomial), we can derive the time dependent dephasing factor based on the theoretical framework previously established in Ref.~\onlinecite{JChemPhys.149.094107}. Here, we further derive the general expression of the time domain dephasing factor for an arbitrary degree of the denominator polynomial $D(p)$.
In general, a real polynomial can have both real and complex roots and due to the fact that its nonreal complex roots always occur in pairs, the denominator polynomial $D(p)$ can be decomposed as
\begin{equation}
\label{eq:denopoly}
   D(p) = \prod_{j=1}^{r}(p-a_{j})^{e_{j}}\prod_{j=1}^{c}[(p-b_{j})(p-b_{j}^{*})]^{\epsilon_{j}},
\end{equation}
where $a_{1},\cdots,a_{r}$ are the $r$ mutually different real roots, $b_{1},\cdots,b_{c}$ are the $c$ mutually different pair of nonreal roots, and we have assumed that the leading coefficient of $D(p)$ is 1.
The positive integer exponents $e_{j}$ and $\epsilon_{j}$ in the decomposition of the denominator polynomial $D(p)$ satisfy $\sum_{j}^{r}e_{j}+2\sum_{j}^{c}\epsilon_{j}=n$.
Consequently, the decomposition of the Laplace domain dephasing factor $\widetilde{F}(p)$ can be expressed as partial fractions
\begin{equation}
\label{eq:polynomials}
\begin{split}
   \widetilde{F}(p)=&\frac{N(p)}{\prod_{j=1}^{r}(p-a_{j})^{e_{j}}\prod_{j=1}^{c}[(p-b_{j})(p-b_{j}^{*})]^{\epsilon_{j}}}\\
                   =&\sum_{j=1}^{r}\Big[\frac{\alpha_{j1}}{(p-a_{j})^{e_{j}}}+\cdots+\frac{\alpha_{je_{1}}}{(p-a_{j})}\Big]\\
                    &+\sum_{j=1}^{c}\Big\{\Big[\frac{\beta_{j1}}{(p-b_{j})^{\epsilon_{j}}}+\frac{\beta_{j1}^{*}}{(p-b_{j}^{*})^{\epsilon_{j}}}\Big]\\
                    &+\cdots+\Big[\frac{\beta_{j\epsilon_{1}}}{(p-b_{j})}+\frac{\beta_{j\epsilon_{1}}^{*}}{(p-b_{j}^{*})}\Big]\Big\},
\end{split}
\end{equation}
where the real coefficients $\alpha_{jk}$ and the complex coefficients $\beta_{jk}$ satisfy
\begin{equation}
\label{eq:coef}
\begin{split}
   \alpha_{jk}&=\frac{1}{(k-1)!}\Big\{\frac{d^{k-1}}{dp^{k-1}}[\widetilde{F}(p)(p-a_{j})^{e_{j}}]\Big\}_{p=a_{j}},\\
   \beta_{jk}&=\frac{1}{(k-1)!}\Big\{\frac{d^{k-1}}{dp^{k-1}}[\widetilde{F}(p)(p-b_{j})^{\epsilon_{j}}]\Big\}_{p=b_{j}}.
\end{split}
\end{equation}
By means of the inverse Laplace transform, the dephasing factor in time domain can be expressed as
\begin{equation}
\label{eq:depfac}
\begin{split}
   F(t) = &\sum_{j=1}^{r}\Big[\frac{\alpha_{j1}t^{e_{j}-1}}{(e_{j}-1)!}+\cdots+\alpha_{je_{1}}\Big]e^{a_{j}t}\\
                   &+\sum_{j=1}^{c}\Big\{\Big[\frac{\beta_{j1}t^{\epsilon_{j}-1}}{(\epsilon_{j}-1)!}+\cdots+\beta_{j\epsilon_{1}}\Big]e^{b_{j}t}\\
                   &+\Big[\frac{\beta_{j1}^{*}t^{\epsilon_{j}-1}}{(\epsilon_{j}-1)!}+\cdots+\beta_{j\epsilon_{1}}^{*}\Big]e^{b_{j}^{*}t}\Big\}.
\end{split}
\end{equation}

Correspondingly, the first derivative of the dephasing factor in Laplace domain satisfies
\begin{equation}
\label{eq:derdepfac}
   \widetilde{\dot{F}}(p) =p\widetilde{F}(p)-F(0)=-\frac{\nu^{2}}{p\big[p+2\lambda\widetilde{K}(p)\big]+\nu^{2}}.
\end{equation}
We can obtain the first time derivative of the dephasing factor by directly taking derivation over the time domain dephasing factor in Eq.~\eqref{eq:depfac} or by means of inverse Laplace transform similarly as we dealt with above.
As a consequence, the dephasing rate can be expressed as
\begin{equation}
\label{eq:depratexp}
   \gamma(t)=-\frac{\dot{F}(t)}{F(t)}
   =-\frac{\mathcal{L}^{-1}\big[\widetilde{\dot{F}}(p)\big]}{\mathcal{L}^{-1}\big[\widetilde{F}(p)\big]}.
\end{equation}
Based on the dephasing rate $\gamma(t)$ and the definition in Eq.~\eqref{eq:non-Mar}, we can further obtain the maximal backflow of coherence information from the environment, namely, the non-Markovianity.
However, to derive the expression of the non-Markovianity, we should know the discrete time constants $t_{1}, t_{2}, \cdots, t_{n}$ for the absolute of the dephasing factor $|F(t)|$ to obtain the relative extrema, namely, the stationary points and singular points by $d/dt|F(t)|=0$.
Due to the continuity of the dephasing factor $F(t)$ and based on the initial condition $F(0)=1$,
the stationary and singular points for the relative minima and maxima in the absolute of the dephasing factor $|F(t)|$ appear alternately.
Consequently, the non-Markovianity can be formally expressed as
\begin{equation}
\label{eq:non-Mar1}
 \mathcal{N}=-\int_{0 \ \gamma(t)<0}^{\infty}\gamma(t)|F(t)|dt=-\sum_{j=1}\int_{t_{2j-1}}^{t_{2j}}\gamma(t)|F(t)|dt,
\end{equation}
where the integral in the sum is taken over all the time intervals $[t_{2j-1}, t_{2j}]$ in which the dephasing rate $\gamma(t)$ is below zero.
In general, the number of the time intervals $[t_{2j-1}, t_{2j}]$ is infinite and it is very difficult to calculate the non-Markovianity analytically by means of the expression in Eq.~\eqref{eq:non-Mar1}.
We here introduce an relatively easy method to calculate the non-Markovianity numerically by rewriting Eq.~\eqref{eq:deprat} as
\begin{equation}
\label{eq:absdepfac}
\frac{d}{dt}|F(t)|=-\gamma(t)|F(t)|,
\end{equation}
By taking an integral over time of Eq.~\eqref{eq:absdepfac} from $t=0$ to $t=\infty$, we obtain the relation as follows
\begin{equation}
\label{eq:inteabsdepfac}
|F(\infty)|-|F(0)|=-\int_{0}^{\infty}\gamma(t)|F(t)|dt.
\end{equation}
Based on the definition for non-Markovianity in Eq.~\eqref{eq:non-Mar} and the vanishing of the dephasing factor $F(\infty)=0$ in the long time limit, the non-Markovianity via numerical calculation can be easily expressed as
\begin{equation}
\label{eq:non-Mar2}
 \mathcal{N}=\frac{1}{2}\left[\int_{0}^{\infty}|\gamma(t)F(t)|dt-1\right].
\end{equation}

\section{Theoretical results and discussion}

There are two dynamics regimes which can be identified in terms of the jumping amplitude $\nu$ and transition rate $\lambda$ of the environmental noise exhibiting only Markovian character~\cite{PhysRevB75.054515,PhysRevB77.174509}:
the weak coupling regime with $\nu<\lambda$ and the strong coupling regime with $\nu>\lambda$.
In the following, we mainly focus on the dephasing dynamics of the quantum system in the two dynamics regimes by considering three important types environmental memory kernel.

\subsection{Memoryless limit}

We first consider the case that environmental noise is in the memoryless limit $K(t-\tau)=\delta(t-\tau)$,
e.g., the statistical property of the environment is Markovian.
In this case, the Laplace transform of the memory kernel yields $\widetilde{K}(p)=1$ and thus the Laplace domain dephasing factor in Eq.~\eqref{eq:depfacexp} and its first derivative in Eq.~\eqref{eq:derdepfac} can be reduced to
\begin{equation}
\label{eq:FlRTN1}
\widetilde{F}(p)=\frac{p+2\lambda}{p(p+2\lambda)+\nu^{2}},
\quad\widetilde{\dot{F}}(p)= -\frac{\nu^{2}}{p(p+2\lambda)+\nu^{2}}.
\end{equation}
By taking the inverse Laplace transform of Eq.~\eqref{eq:FlRTN1}, we can obtain the dephasing factor in time domain as
\begin{equation}
\label{eq:FtRTN}
F(t)= e^{-\lambda t}
\begin{cases}
\cosh\big(t\sqrt{\lambda^{2}-\nu^{2}}\big)+\frac{\lambda}{\sqrt{\lambda^{2}-\nu^{2}}}\sinh\big(t\sqrt{\lambda^{2}-\nu^{2}} \big),\\
1+\lambda t,\\
\cos\big(t\sqrt{\nu^{2}-\lambda^{2}}\big)+\frac{\lambda}{\sqrt{\nu^{2}-\lambda^{2}}}\sin\big(t\sqrt{\nu^{2}-\lambda^{2}}\big),
\end{cases}
\end{equation}
for $\nu<\lambda$, $\nu=\lambda$ and $\nu>\lambda$, respectively.
This expression is consistent with the well-known results obtained in Refs.~\onlinecite{PhysRevB78.201302,PhysStatusSolidiB246.1018,PhysRevA89.012330}.
Correspondingly, the dephasing rate can be written as
\begin{equation}
\label{eq:gammat}
\gamma(t)=
\begin{cases}
\frac{\nu^{2}\sinh\big(t\sqrt{\lambda^{2}-\nu^{2}}\big) }{\sqrt{\lambda^{2}-\nu^{2}}\cosh\big(t\sqrt{\lambda^{2}-\nu^{2}}\big)+\lambda\sinh\big(t\sqrt{\lambda^{2}-\nu^{2}}\big)}
,&\nu<\lambda,\\
\frac{\lambda^{2}t}{1+\lambda t},&\nu=\lambda,\\
\frac{\nu^{2}\sin\big(t\sqrt{\nu^{2}-\lambda^{2}}\big) }{\sqrt{\nu^{2}-\lambda^{2}}\cos\big(t\sqrt{\nu^{2}-\lambda^{2}}\big)+\lambda\sin\big(t\sqrt{\nu^{2}-\lambda^{2}}\big)},&\nu>\lambda.
\end{cases}
\end{equation}

\begin{figure*}[ht!]
 \centering
    \includegraphics[width=5.0in]{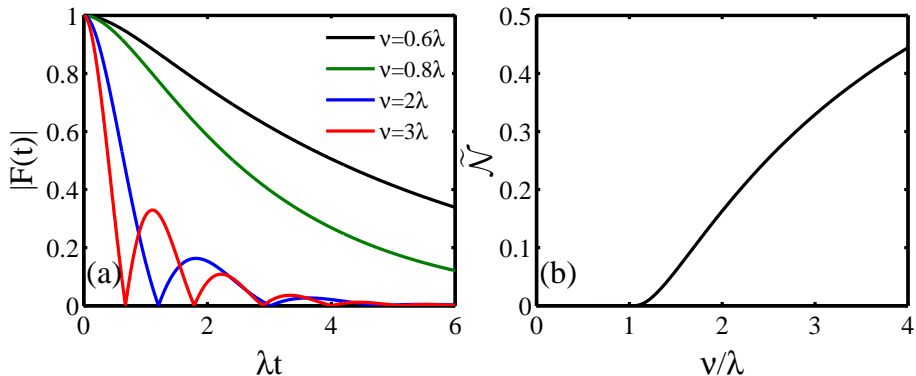}
    \caption{(Color online) (a)~Time evolution of the dephasing factor $|F(t)|$ induced by the Markovian RTN
    for different values of the coupling $\nu$. (b)~Scaled non-Markovianity $\widetilde{\mathcal{N}}=\mathcal{N}/(\mathcal{N}+1))$ as a function of the coupling $\nu$.}
    \label{fig:MRTNFN}
\end{figure*}

Figure~\ref{fig:MRTNFN} (a) shows the dephasing factor $|F(t)|$ as a function of the evolution time induced by the RTN in the memoryless limit for different values of the coupling $\nu$.
In weak coupling ($\nu<\lambda$), the dephasing factor only displays monotonical decay which reflects that the quantum dephasing dynamics of the system is always Markovian and the non-Markovianity is zero.
Furthermore, the decay in the dephasing factor become fast as the coupling increases implying
that the coupling can enhance the quantum dephasing dynamics of the system.
In strong coupling ($\nu>\lambda$), the decay in the dephasing factor is nonmonotonical with periodic zeros which indicates that the quantum dephasing dynamics of the system is non-Markovian and the non-Markovianity is no longer zero due to coherence revivals. In addition, as the coupling increases, the oscillatory behavior exhibited in the dephasing factor is more and more obvious and the emergence of coherence revivals becomes ahead of time.
Figure~\ref{fig:MRTNFN} (b) shows the non-Markovianity $\mathcal{N}$ as a function of the coupling $\nu$.
The quantum dynamical dephasing of the system displays a transition from Markovian (zero non-Markovianity) to non-Markovian (non-zero non-Markovianity) as the coupling increases with the boundary of the weak and strong coupling regimes $\nu=\lambda$.
In the strong coupling regime, the non-Markovianity become larger as the coupling increases which suggests the enhancement of the non-Markovian character in the quantum dephasing dynamics of the system.

\subsection{Exponential memory kernel}

We consider that the statistical property of the environmental noise is non-Markovian,
where the memory kernel of the environment is of a widely used exponential form
$K(t-\tau)=\kappa e^{-\kappa(t-\tau)}$ with the decay rate $\kappa$~\cite{PhysRevA73.012111,PhysRevA81.062120}.
In this case, the Laplace transform of the memory kernel satisfies $\widetilde{K}(p)=\frac{\kappa}{p+\kappa}$. Thus, in Laplace domain the dephasing factor together with its first derivative in Eqs.~\eqref{eq:depfacexp} and~\eqref{eq:derdepfac} can be respectively expressed as
\begin{equation}
\label{eq:FlRTN2}
\widetilde{F}(p)=\frac{p+\frac{2\lambda\kappa}{p+\kappa}}{p\left(p+\frac{2\lambda\kappa}{p+\kappa}\right)+\nu^{2}},
\quad\widetilde{\dot{F}}(p)= -\frac{\nu^{2}}{p\left(p+\frac{2\lambda\kappa}{p+\kappa}\right)+\nu^{2}}.
\end{equation}
This result is completely compatible with that we obtained in Refs.~\onlinecite{PhysRevA94.042110,JChemPhys.149.094107} when the environmental noise is stationary.
The expression of the dephasing factor in time domain can be, by taking the inverse
Laplace transform of Eq.~\eqref{eq:FlRTN2}, written as the form as we derived in Eq.~\eqref{eq:depfac}
which we also derived in Ref.~\onlinecite{JChemPhys.149.094107}.

\begin{figure*}[ht]
 \centering
    \includegraphics[width=5.0in]{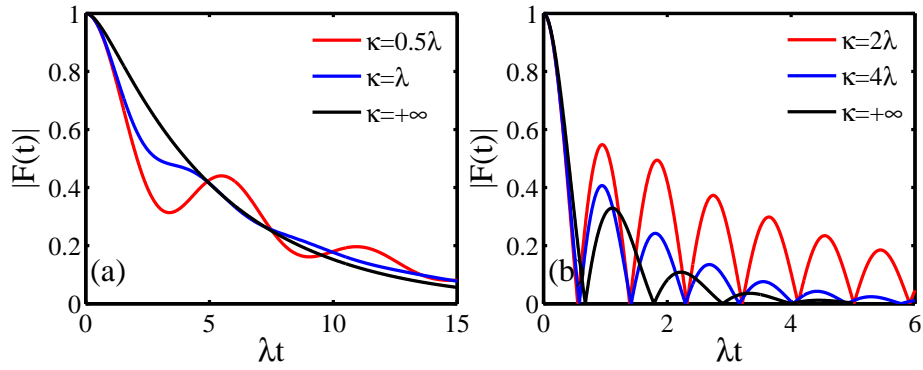}
    \caption{(Color online) Time evolution of the dephasing factor $|F(t)|$ induced by the non-Markovian RTN
    for different values of the memory decay rate $\kappa$ in (a)~the weak coupling regime with $\nu=0.6\lambda$ and (b)~the strong coupling regime with $\nu=3\lambda$.
    Dash black line indicates the memoryless case of the RTN.}
    \label{fig:NMRTNF1}
\end{figure*}

Figure~\ref{fig:NMRTNF1} shows the dephasing factor $|F(t)|$ as a function of the evolution time induced by the
non-Markovian RTN for different values of the memory decay rate $\kappa$.
As shown in Fig.~\ref{fig:NMRTNF1} (a), in the weak coupling regime, the dephasing factor displays nonmonotonical decay with nonzero coherence revivals for a given memory decay rate.
In contrast, in the strong coupling regime, as displayed in Fig.~\ref{fig:NMRTNF1} (b), the dephasing factor shows nonmonotonical oscillatory decay with zero coherence revivals.
In addition, the dephasing factor gets reduced and the nonmonotonical oscillations in the dephasing factor get enhanced as the environmental memory effect increases.
These results imply that the memory effect of the environmental noise can reduce the quantum dynamical dephasing but enhance the non-Markovian character exhibited in the quantum dephasing dynamics of the system.

\begin{figure}[ht]
 \centering
    \includegraphics[width=3.0in]{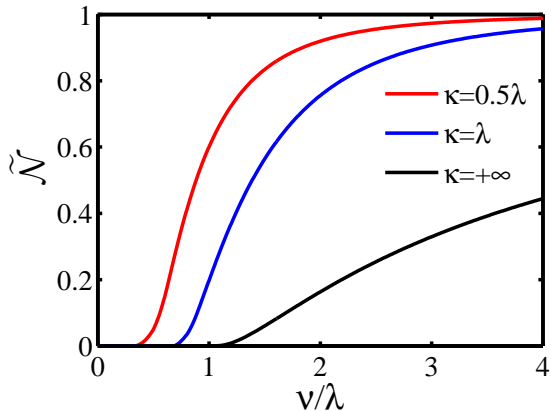}
    \caption{(Color online) Scaled non-Markovianity $\widetilde{\mathcal{N}}=\mathcal{N}/(\mathcal{N}+1)$
    as a function of the coupling $\nu$ for different memory decay rate $\kappa$.
    Dash black line indicates the case of the Markovian RTN.}
    \label{fig:NMRTNNM1}
\end{figure}

Figure~\ref{fig:NMRTNNM1} shows the non-Markovianity $\mathcal{N}$ induced by the non-Markovian RTN as a function of the coupling $\nu$ for different values of the memory decay rate $\kappa$.
For a given memory decay rate, the quantum dephasing dynamics of the system displays an obvious transition from Markovian (zero non-Markovianity) to non-Markovian (nonzero non-Markovianity) as the coupling increases and the transition boundary depends on the memory decay rate: The stronger the environmental memory effect is, the smaller the transition boundary of the coupling is.
In addition, for a given coupling, as the memory decay rate decreases, the non-Markovianity increases.
These results indicate that the non-Markovian character of the environment can increase the non-Markovian character in the  quantum dephasing dynamics of the system and expand the non-Markovian dynamics region.

\subsection{Modulatable memory kernel}

We consider that the environmental noise is a modulatable non-Markovian process,
where the memory kernel of the environment satisfies
$K(t-\tau)=\kappa e^{-\kappa(t-\tau)}\cos[\Omega(t-\tau)]$ with the decay rate $\kappa$
and external modulation frequency $\Omega$ of the environment~\cite{PhysRevA30.568,PhysRevE83.041104}.
In this case, the memory kernel in Laplace domain yields $\widetilde{K}(p)=\frac{\kappa(p+\kappa)}{(p+\kappa)^{2}+\Omega^{2}}$
and the dephasing factor in Eq.~\eqref{eq:depfacexp} and its first derivative in Eq.~\eqref{eq:derdepfac} can be respectively written as
\begin{equation}
\label{eq:FlRTN3}
\begin{split}
\widetilde{F}(p)&=\frac{p+\frac{2\lambda\kappa(p+\kappa)}{(p+\kappa)^{2}+\Omega^{2}}}{p\left[p+\frac{2\lambda\kappa(p+\kappa)}{(p+\kappa)^{2}+\Omega^{2}}\right]+\nu^{2}},\\
\quad\widetilde{\dot{F}}(p)&= -\frac{\nu^{2}}{p\left[p+\frac{2\lambda\kappa(p+\kappa)}{(p+\kappa)^{2}+\Omega^{2}}\right]+\nu^{2}}.
\end{split}
\end{equation}
The dephasing factor in time domain can be, by taking the inverse
Laplace transform of Eq.~\eqref{eq:FlRTN2}, expressed in the form as we derived in Eq.~\eqref{eq:depfac}.
The expression of the dephasing factor in Eq.~\eqref{eq:FlRTN3} is consistent with that in Eq.~\eqref{eq:FlRTN2} when there is no environmental external modulation $\Omega=0$.

\begin{figure}[ht]
 \centering
    \includegraphics[width=3.5in]{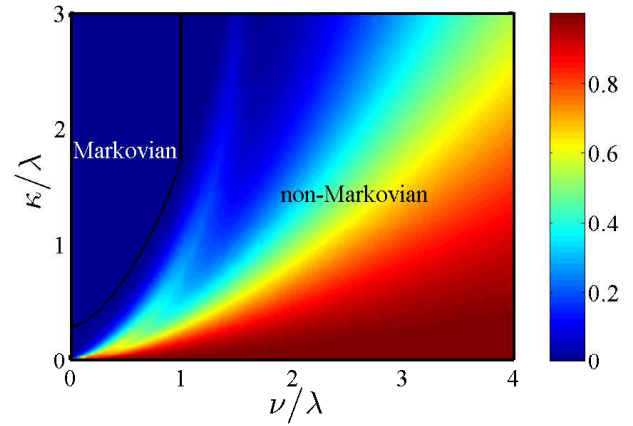}
    \caption{(Color online) Phase diagram via intensity plot of the scaled non-Markovianity $\widetilde{\mathcal{N}}=\mathcal{N}/(\mathcal{N}+1)$ in the $\kappa\sim\nu$ plane for the transition between Markovian and non-Markovian dephasing dynamics with no frequency modulation $\Omega=0$.}
    \label{Figphase}
\end{figure}

We first plot the phase diagram in the $\kappa\sim\nu$ plane to identify Markovian and non-Markovian quantum dynamics regions of the system in the presence of the non-Markovian RTN without frequency modulation $\Omega=0$ in Fig.~\ref{Figphase}.
As shown in the figure, we can clearly see that the quantum dynamics of the system is not always non-Markovian when the environmental statistical property is non-Markovian, and that the non-Markovian character in the quantum dephasing dynamics depends both on the environmental non-Markovian character and the coupling between the system and the environment~\cite{PhysRevA94.042110}.
In strong coupling ($\nu>\lambda$), the quantum dephasing dynamics of the system is always non-Markovian (nonzero non-Markovianity) whereas it exhibits a transition from non-Markovian (nonzero non-Markovianity) to Markovian (zero non-Markovianity) in weak coupling ($\nu<\lambda$) resulting from the memory effect of the environmental noise: The stronger the coupling is (the larger $\nu$ is), the larger the threshold value of the memory decay rate $\kappa_{\mathrm{th}}$ for the transition boundary is. In addition, the threshold value $\kappa_{\mathrm{th}}$ depends closely on the value of the coupling $\nu$. For example, the threshold value of the memory decay rate $\kappa$ is $\kappa_{\mathrm{th}}=1.23\lambda$ for the coupling $\nu=0.8\lambda$.

\begin{figure*}[ht]
 \centering
    \includegraphics[width=5.0in]{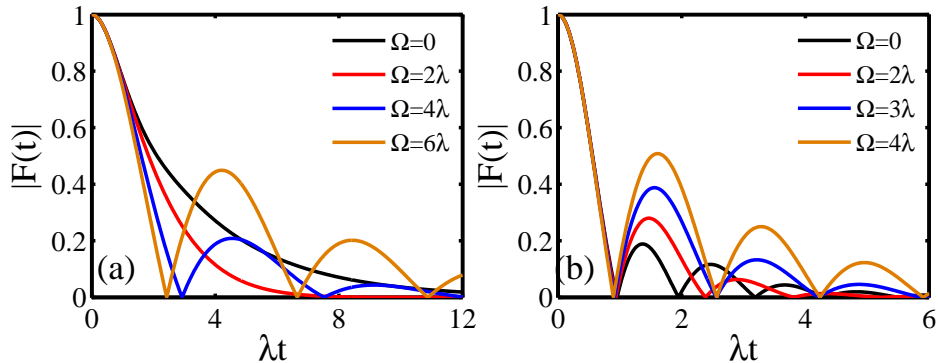}
    \caption{(Color online) Time evolution of the dephasing factor $|F(t)|$ induced by the non-Markovian RTN
    in two dynamics regions identified in Fig.~\ref{Figphase} for different values of the external modulation frequency $\Omega$ of the environment (a)~the Markovian region with $\nu=0.8\lambda$ and (b)~the non-Markovian region with $\nu=3\lambda$. The memory decay rate $\kappa=3\lambda$.
    Dash black line indicates the case with no frequency modulation $\Omega=0$.}
    \label{fig:NMRTNF2}
\end{figure*}

Figure~\ref{fig:NMRTNF2} shows the time evolution of the dephasing factor $|F(t)|$ induced by the non-Markovian RTN for different values of the external modulation frequency $\Omega$ of the environment in two dynamics regions identified in Fig.~\ref{Figphase}.
In the Markovian dynamics region as shown in Fig.~\ref{fig:NMRTNF2} (a), the dephasing factor first decays fast and then starts to display non-Markovian character as the external modulation frequency of the environment increases.
These results indicate that the environmental external modulation can make the quantum dephasing dynamics of the system undergo a
transition between Markovian (zero non-Markovianity) and non-Markovian (nonzero non-Markovianity) in this dynamics region.
In contrast to the Markovian dynamics region, the environmental external modulation can only enhance the non-Markovian character in the quantum dephasing dynamics in non-Markovian dynamics region as displayed in Fig.~\ref{fig:NMRTNF2} (b).
These results suggest that the quantum dephasing dynamics can be well modulated by the external modulation frequency of the environment.

\section{Conclusions}
\label{sec:Conc}
We have theoretically studied the dynamical dephasing of a quantum system coupled to an environment exhibiting non-Markovian random telegraph fluctuations depending on the environmental memory effect.
We have derived the exact expression of the dephasing factor closely associated with the memory kernel of the environmental noise.
Based on three important types memory kernel of the environmental noise, we have shown that the dephasing dynamics of the quantum system is not always non-Markovian, and that the non-Markovian character in the quantum dephasing dynamics depends both on the environmental non-Markovian character and the coupling between the system and environment.
Moreover, we have shown that the quantum dephasing dynamics of the system can be modulated by the external modulation frequency of the environment.
We hope that the investigation in the paper will be helpful for further understanding of the non-Markovian quantum dephasing dynamics of open quantum systems and will be effective in suppressing and controlling the quantum dynamical dephasing in the presence of a non-Markovian environment.

\begin{acknowledgments}
  This work was supported by the National Natural Science Foundation of China (Grant No.11947033).
X.C. also acknowledges the support from the Doctoral Research Fund of Shandong Jianzhu University (Grant No. XNBS1852).
\end{acknowledgments}

%

\end{document}